\documentclass[twocolumn,tightenlines,showkeys,preprintnumbers,floatfix,amsmath,amssymb,superscriptaddress,prl]{revtex4-1}
\usepackage[T1]{fontenc}
\usepackage[utf8]{inputenc}
\usepackage[french, english]{babel}
\usepackage{dcolumn}
\usepackage{bm}
\usepackage{hyperref}
\usepackage[dvips]{graphicx}
\usepackage{hyperref}
\begin{document}

\title{Modulating the spin transfer torque switching dynamics with two orthogonal spin-polarizers by varying the cell aspect ratio}
\author{B. Lacoste}
	\affiliation{SPINTEC, UMR CEA/CNRS/UJF-Grenoble 1/ Grenoble-INP, INAC, Grenoble, F-38054, France}
\author{M. Marins de Castro}
	\affiliation{SPINTEC, UMR CEA/CNRS/UJF-Grenoble 1/ Grenoble-INP, INAC, Grenoble, F-38054, France}
\author{T. Devolder}
	\affiliation{Institut d'Electronique Fondamentale, CNRS UMR 8622, Bat. 220, Universit\'{e} Paris-Sud, 91405 Orsay, France}
\author{R. C. Sousa}
	\affiliation{SPINTEC, UMR CEA/CNRS/UJF-Grenoble 1/ Grenoble-INP, INAC, Grenoble, F-38054, France}
\author{L. D. Buda-Prejbeanu}
	\affiliation{SPINTEC, UMR CEA/CNRS/UJF-Grenoble 1/ Grenoble-INP, INAC, Grenoble, F-38054, France}
\author{S. Auffret}
	\affiliation{SPINTEC, UMR CEA/CNRS/UJF-Grenoble 1/ Grenoble-INP, INAC, Grenoble, F-38054, France}
\author{U. Ebels}
	\affiliation{SPINTEC, UMR CEA/CNRS/UJF-Grenoble 1/ Grenoble-INP, INAC, Grenoble, F-38054, France}
\author{C. Ducruet}
	\affiliation{Crocus Technology, 38025 Grenoble, France}
\author{I. L. Prejbeanu}
	\affiliation{Crocus Technology, 38025 Grenoble, France}
\author{L. Vila}
	\affiliation{SP2M/NM, CEA/Grenoble, INAC, 38054 Grenoble Cedex, France}
\author{B. Rodmacq}
	\affiliation{SPINTEC, UMR CEA/CNRS/UJF-Grenoble 1/ Grenoble-INP, INAC, Grenoble, F-38054, France}
\author{B. Dieny}
	\affiliation{SPINTEC, UMR CEA/CNRS/UJF-Grenoble 1/ Grenoble-INP, INAC, Grenoble, F-38054, France}
\date{Date of submission \today}

\begin{abstract}
We study in-plane magnetic tunnel junctions with additional perpendicular polarizer for subnanosecond current induced switching memories. The spin-transfer-torque switching dynamics was studied as a function of the cell aspect ratio both experimentally and by numerical simulations using the macrospin model. We show that the anisotropy field plays a significant role in the dynamics, along with the relative amplitude of the two spin-torque contributions. This was confirmed by micromagnetic simulations. Real-time measurements of the reversal were performed with samples of low and high aspect ratio. For low aspect ratio, a precessional motion of the magnetization was observed and the effect of temperature on the precession coherence was studied. For high aspect ratio, we observed magnetization reversals in less than 1~ns for large enough current densities, the final state being controlled by the current direction in the MTJ cell.
\end{abstract}

\maketitle

Spin-transfer torque magnetic random-access memory (STT-MRAM) are very promising non-volatile memories envisioned to provide devices with smaller sizes and faster dynamics. A conventional STT-MRAM consists of a reference layer, which magnetization is fixed either in-plane or out-of-plane, separated by an MgO barrier from the storage layer (SL), which magnetization is free. The storage layer magnetization has two stable configurations, parallel (P) or antiparallel (AP) to the reference layer. To write the memory cell, a voltage pulse is applied to the magnetic tunnel junction (MTJ) that can reverse the storage layer magnetization, thanks to the spin-transfer torque (STT) of the polarized current due to the reference layer. However the STT is proportional to the vectorial product of the reference layer magnetization with the storage layer magnetization, so that, in the equilibrium configuration (P or AP), the STT vanishes. The reversal of the MTJ is only possible thanks to thermal fluctuations that misalign the two layer magnetizations resulting in a stochastic switching dynamics in conventional STT-MRAM. In fact, even if the switching itself lasts less than a~nanosecond, the switching occurs after a random incubation time. This is detrimental to the switching time, as it is difficult to switch an MTJ with a bit error rate lower than $10^{-4}$ in less than 10~ns~\cite{Devolder2008, Koch2004}, which is necessary for application as a fast RAM.\par
In order to eliminate the incubation time, it was proposed to add another polarizing layer with a magnetization fixed and orthogonal to the equilibrium directions of the SL magnetization, in order to maximize the STT acting on the storage layer magnetization as soon as the write current pulse is switched on, while the SL magnetization is still aligned along its equilibrium direction~\cite{Redon2001, Lee2009, MarinsdeCastro2012, Kent2004}. Switching times below 1~ns were observed with this design~\cite{Liu2010}. The same configuration is studied here. The in-plane storage layer is separated from the in-plane reference layer by an MgO barrier, as compared to previous work with a spin valve~\cite{Houssameddine2007}, to improve the spin transfer torque from the reference layer and to obtain an output signal large enough for time-resolved measurements. An additional perpendicular polarizer (PP), whose magnetization is out-of-plane, is separated from the storage layer by a non-magnetic spacer. Although it was previously observed that the presence of the PP reduces switching times, the magnetization dynamics with two polarizing layers is not completely understood, especially the relative influence of the two polarizers. Here we propose a theoretical model, confirmed experimentally, to describe the cross-over between precessional motion of the SL magnetization and switching with two polarizers, due to a change in the anisotropy field.\\
\par

\begin{figure}[b]
\centering
\includegraphics[width=0.9\linewidth]{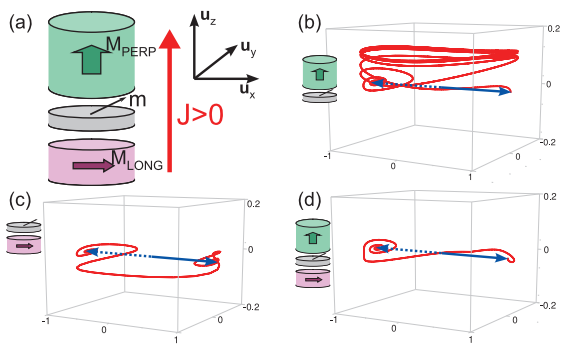}
\caption{(a) Geometry of the nanopillars with elliptical cross-section. (b)-(d) Magnetization dynamics during the reversal of the storage layer in three configurations~: (b) with a perpendicular polarizer only, (c) with an in-plane reference layer only and (d) with the two polarizing layers. The magnetization is initially in the parallel (P) configuration (blue full-arrow), and relaxes in the anti-parallel (AP) state (blue dotted-arrow).}
\label{fig:traj}
\end{figure}
 
With this geometry, described in Fig.~\ref{fig:traj}.(a), the SL magnetization is submitted to two STT contributions, which have different effects on the SL magnetization dynamics. Notice that this analysis can be applied to the modeling of a free layer with spin orbit torque in arbitrary direction, because the macrospin equation is similar~\cite{Park2014} or for out-of-plane MTJ with additional in-plane polarizer.
In the following, $M_s$ is the saturation magnetization of the storage layer, $H_K$ the in-plane anisotropy field, $M_{\operatorname{eff}}$ the reduced demagnetizing field (due to interface perpendicular anisotropy for instance), $t$ the thickness of the storage layer, $\alpha$ the Gilbert damping constant, $ \eta_{\operatorname{LONG}} $ and $ \eta_{\operatorname{PERP}} $ the STT efficiency of the reference layer and of the perpendicular polarizer.\par

On the one hand, the STT contribution from the perpendicular polarizer pulls the SL magnetization out-of-plane [Fig.~\ref{fig:traj}.(b)], then due to the strong demagnetizing field, the FL magnetization precesses around the out-of-plane axis at gigahertz frequency. The free layer magnetization is in out-of-plane precession (OPP) around the $z$-axis when a current density larger than the critical current density $J^{\operatorname{PERP}}_{c}$~\cite{Lee2005, Ebels2008} is applied~:
\begin{equation}
J^{\operatorname{PERP}}_{c}=\frac {2e}{\hbar}\frac {\mu_0 M_{s} t}{\eta_{\operatorname{PERP}}}\frac {H_{K}}{2}
\label{eq:Jperp}
\end{equation}
By tailoring the current pulse width, it is possible to stop the FL magnetization precession after half a precession, hence reversing the magnetization direction and switching the device~\cite{MarinsdeCastro2012,Kent2004}. However, achieving a 180$^{\circ}$ SL magnetization rotation implies being able to control the pulse duration with a typical accuracy of 200ps~$\pm$50ps~\cite{Papusoi2009}. This is possible at the single cell level but much more difficult at a memory chip level due to the deformation of the current pulses during their propagation along the bit lines. Furthermore, since this write procedure is similar to a toggle writing, it requires to read before write.
\par
On the other hand, the STT contribution from the analyzer provokes a bipolar switching of the FL magnetization [Fig.~\ref{fig:traj}.(c)]. The expression of the critical current density $J^{\operatorname{LONG}}_{c}$ comes out from the study of the equilibrium stability~\cite{Slonczewski1996,Sun2000,Li2003}:
\begin{equation}
J^{\operatorname{LONG}}_{c}=\frac {2e}{\hbar}\frac {\mu_0 M_{s} t}{\eta_{\operatorname{LONG}}}\alpha(\dfrac{M_{\operatorname{eff}}}{2}+H_{K})
\label{eq:Jlong}
\end{equation}
Depending on the polarity of the current, one of the two stable configurations, parallel (P) or antiparallel (AP) is favored, so that the final written state can be controlled by the current direction. However the switching is then stochastic as previously explained.\par
By combining the two STT contributions from the two orthogonal polarizing layers, one can expect to be able to still control the final state by the current pulse direction through the MTJ while reducing the stochasticity of the switching thanks to the STT contribution from the PP [Fig.~\ref{fig:traj}.(d)]. But this requires to properly tune the relative amplitude of these two STT contributions~\cite{Hou2011,Mejdoubi2013,Park2013,Liu2010,Bazaliy2012}.\\
However, from the expressions of the two critical currents, it appears that if the uniaxial anisotropy field $H_K$ is increased, the critical current that controls the appearance of the precessional motion is also increased (Eq.~\ref{eq:Jperp}), while the critical current for bipolar switching is not changed much because $H_K \ll M_{\operatorname{eff}}$ for in-plane MTJ. Hence, instead of tuning the relative amplitude of the two STT contributions, this qualitative analysis suggests to increase the anisotropy field $H_K$ to favor the bipolar switching over the precessional regime.\\

\par
If the two STT contributions of the reference layer and of the PP are included in the Landau-Lifshitz-Gilbert-Slonczewski equation that describes the dynamics of the free layer, the equilibrium analysis in the macrospin approximation exhibits two critical current densities, which values are close to $J^{\operatorname{LONG}}_{c}$ and $J^{\operatorname{PERP}}_{c}$. A more extended calculation of the critical currents is presented in Appendix~\ref{Annex-1}. In a nutshell, below $J_c^{\operatorname{LONG}}$, the magnetization remains in equilibrium. And above $J_c^{\operatorname{PERP}}$, only the dynamic OPP state exists. However, between these two current densities, both the OPP and switched state can be reached. However, the critical current $J^{\operatorname{OPP}}_{c}$ below which the OPP state cannot exist was computed by studying the stability of the OPP with an anisotropy field and the STT from the reference layer~\cite{Lacoste2013}~:
\begin{equation}
J^{\operatorname{OPP}}_{c}=\frac {2e}{\hbar}\frac {\mu_0 M_{s} t}{\eta_{\operatorname{PERP}}}\dfrac{\alpha}{2}\sqrt{H_K M_{\operatorname{eff}}}
\label{eq:JOPP}
\end{equation}
This critical current depends on the anisotropy field $H_K$, but because of the square root dependence it is smaller than $J^{\operatorname{PERP}}_{c}$. Between these two critical current densities, the magnetization can be in two bistable states: OPP or switched state. The final state depends on the detail of the dynamics.\\

\begin{figure}[t]
\centering
\includegraphics[width=0.8\linewidth]{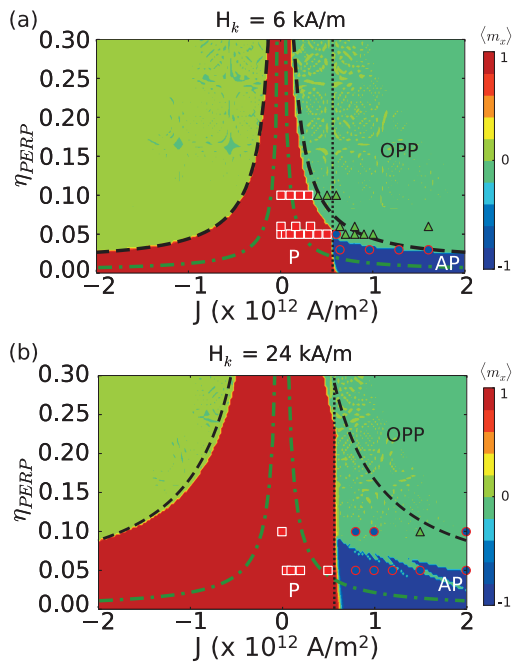}
\caption{Macrospin simulation of the average in-plane magnetization $m_x$, versus current density and polarization of the perpendicular polarizer, with $\eta_{\operatorname{LONG}}=0.3$, for different anisotropy field (a) 6~kA/m and (b) 24~kA/m. There are three final steady states~: (i) in red, P final state, no switching. (ii) in blue, AP final state, switching of the SL magnetization. (iii) in green, the average in-plane component of the magnetization vanishes, corresponding to an OPP steady state. Black dotted-line at the right of which the P equilibrium is unstable. Black dashed-line above which only OPP exist. Green dash-dotted-line below which OPP cannot exist. Symbols represent the final steady state from micromagnetic simulations~: P final state (orange squares), AP final state (blue circles), and OPP (green triangles).}
\label{fig:macrospin}
\end{figure}

To describe the bistable region AP/OPP and to validate the critical line expressions, we performed macrospin simulations with different anisotropy field $H_k$ and polarization of the perpendicular polarizer $\eta_{\operatorname{PERP}}$. The parameters for the simulations are~: $\alpha=0.02$, $t=3$~nm, $M_S=M_{\operatorname{eff}}=1.2\times10^6$~A/m, $\eta_{\operatorname{LONG}}=0.3$ and the magnetization is initially in the P state ($m_x=1$). The average in-plane magnetization component $m_x$ in the permanent regime is calculated for different values of applied current density $J_{\operatorname{app}}$ and polarization $\eta_{\operatorname{PERP}}$ and represented in Fig.~\ref{fig:macrospin}, for (a) $H_k=6\:$kA/m and (b) $H_k=24\:$kA/m. The diagrams show three regions~: (i) in red, the final state remains the initial P state, the SL has not switched. (ii) in blue, the final state is the AP state, the SL has switched. (iii) in green, the SL is in OPP steady state, the final state depends on the current density pulse duration. The analytical critical line for reversal, in black dotted line, is in agreement with the macrospin simulations. However, the border between the OPP (green) region and the switching (blue) region does not correspond to any theoretical critical line, as it stands in the bistable region delimited by the black dashed-line, of appearance of OPP, and the green dash-dotted-line, of disappearance of OPP. Notice that for negative current densities, the black dashed critical line is in agreement with the simulations, because the initial P equilibrium is stable until this critical line. The effect of the anisotropy field was confirmed by the simulations~: if the anisotropy field is increased, the range of bipolar switching (blue region) is increased, at the expense of the OPP (red) region.
\par
We also confirmed the impact of the anisotropy on the reversal with micromagnetic simulations with two polarizing layers. The final state after 10~ns is reported in Fig.~\ref{fig:macrospin} by the symbols. The micromagnetic simulations were carried out on a cylindrical free layer with elliptical section of dimensions $105\times95\times3$~nm, that corresponds to $H_k=6\:$kA/m, and $180\times60\times3$~nm, to $H_k=24\:$kA/m. The exchange stiffness constant was set to 1.6$\times 10^{-11}$~J/m, all the other parameters being the same as for the macrospin simulations.\\

\begin{figure}[h]
\centering
\includegraphics[width=0.9\linewidth]{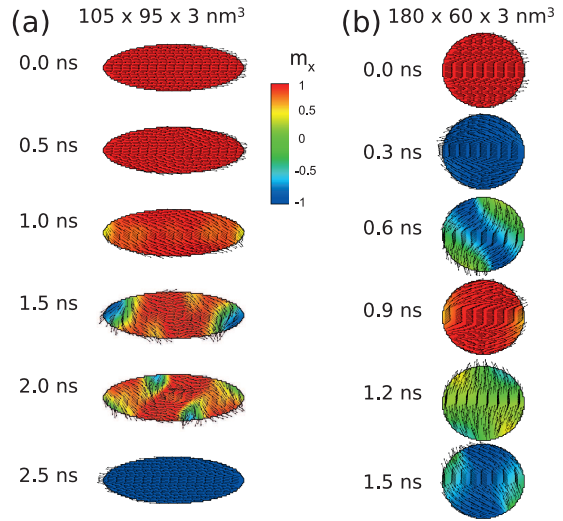}
\caption{Snapshots of the micromagnetic configuration of the free layer at different times for (a) a large aspect ratio of $180\times60\times3$~nm and (b) a small aspect ratio of $105\times95\times3$~nm. The spin polarization of the reference layer and the perpendicular polarizer are $\eta_{LONG}=0.3$ and $\eta_{PERP}=0.05$ respectively. The current density is $10^{12}$~A/m$^2$.}
\label{fig:micromag}
\end{figure}

In Fig.~\ref{fig:macrospin}, the boundary for the stability of the initial P configuration is similar in micromagnetics than in the macrospin model: the orange squares, that stand for a final P state not reversed, are situated in the P region. This is due to the fact that the initial micromagnetic configuration is uniformly magnetized, except at the edges, very similar to the macrospin approximation. However, the boundary between the reversal and the precessional state differs in micromagnetic simulations. It appears that the OPP state is less stable in micromagnetics, mainly because the precession is not spatially uniform, so the macrospin picture is not valid anymore. As a result, in the bistable AP/OPP region, some set of parameters for which a final OPP state was observed in macrospin appear to be reversals in micromagnetic simulations. For large current densities and large perpendicular polarizer spin polarization $\eta_{PERP}$, though, a non-uniform, large amplitude out-of-plane precessional motion was observed, in agreement with the macrospin analysis.\\
As for the effect of the aspect ratio, the trend is the same as predicted by the macrospin simulations, larger aspect ratio favor switching. Fig.~\ref{fig:micromag} shows snapshots of the micromagnetic configuration at different times for a high aspect ratio and a low aspect ratio, with a current density of $10^{12}$~A/m$^2$. For a large aspect ratio, a reversal of the free layer magnetization is observed, whereas with a low aspect ratio, and the other parameters kept unchanged, we found a large amplitude oscillations dynamical state. This is in agreement with the macrospin study.\\

\par
\begin{figure}[t]
\centering
\includegraphics[width=\linewidth]{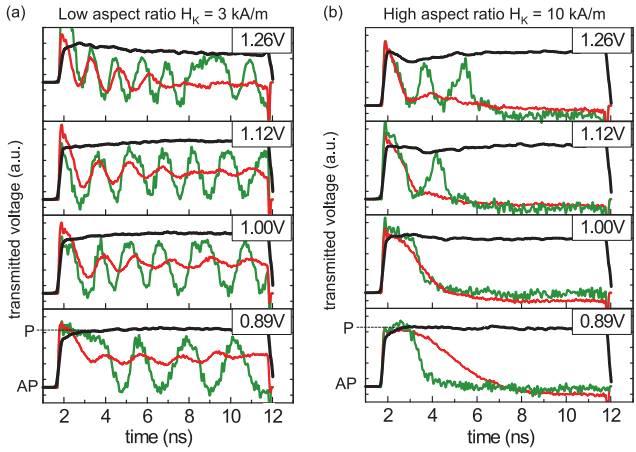}
\caption{Transmitted voltage with an applied voltage pulse of 0.89~V, 1.00~V, 1.12~V and 1.26~V (from bottom to top) in a MTJ nanopillar with (a) low aspect ratio and (b) high aspect ratio. The MTJ is initially in P state. In black, reference resistance in the AP state taken with an external field to saturate the junction. In green, a single-shot trace. In red, average of 50 traces.}
\label{fig:exp-both-SS}
\end{figure}

In order to demonstrate the impact of the aspect ratio, in-plane MTJ stacks with perpendicular polarizer were grown and patterned with different aspect ratios. Real-time measurements of the resistance change were then performed on these samples submitted to voltage pulses. The stacks are described in Ref.~\cite{MarinsdeCastro2012}. They consist of, from bottom to top~: a synthetic antiferromagnet perpendicular polarizer /3~nm Cu spacer/free (storage) layer/MgO barrier/reference layer. The perpendicular polarizer is a synthetic antiferromagnetic multilayer of composition Ta~3/Pt 5/[Co~0.5/Pt~0.4]x5/Co 0.5/Ru 0.9/[Co~0.5/Pt~0.4]x3/Co 0.5/CoFeB~1~nm. The free layer is also a synthetic antiferromagnetic stack consisting of CoFeB~1.3/Ru~0.9/CoFeB~1.7~nm. The reference layer is made of CoFeB~3/Ru~0.9/Co~2/IrMn~7~nm. The MgO barrier between the storage and reference layer is realized by first depositing Mg, and then by a 10~s natural oxidation under a 160~mbar oxygen pressure.\\
All the layers are synthetic antiferromagnets to minimize their mutual magnetostatic interactions. After deposition, the samples were annealed at 300$^{\circ}\mathrm{C}$ for 90~min under an in-plane magnetic field of 0.23~T. Then the sample was patterned in elliptical nanopillars of various aspect ratios. We measured an average TMR signal of about 70\% and a $R\times A$ product of 17~$\Omega \cdot \mu m^2$. Due to a residual stray field, the antiparallel (AP) alignment is favored in the samples.
\par

The nanopillars are connected to a resistance versus field measurement bench. On top of this setup, at any given field, it is possible to send a voltage pulse of 10~ns width through the nanopillar and measure the transmitted voltage with an oscilloscope in real-time~\cite{Devolder2008}. An external bias field is applied to compensate for the residual stray field on the SL.\\
The switching probabilities versus pulse duration were also measured, by measuring the resistance after the current pulse application and comparing it with the resistance before. These measurements were averaged over one hundred hysteresis curves, by sending the pulse in the center of the hysteresis loop and for different pulse width, ranging from 100~ps to 10~ns.\\
\par


\begin{figure}[t]
\centering
\includegraphics[width=\linewidth]{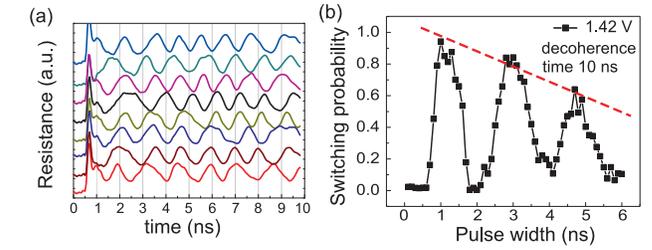}
\caption{Low aspect ratio MTJ at room temperature. (a) Single-shot traces for the same applied voltage pulse of 1.42~V. A low-pass filter at 3~GHz was applied. (b) Switching probability of an MTJ initially in P state versus applied pulse width, with a voltage amplitude set to 1.42~V. Oscillation decay in red dashed line.}
\label{fig:exp-lAR-dec}
\end{figure}

First, we focus on samples with a low aspect ratio of 2.5:1, with nominal sizes of $170\times70\times3$~nm. The in-plane anisotropy field is measured to be around 3~kA/m. In these samples the effect of the PP is dominant, so a precessional motion of the SL magnetization around the out-of-plane axis is expected. Fig.~\ref{fig:exp-both-SS}.(a) shows the transmitted voltage during a pulse of 10~ns and of different voltages, 0.89V, 1.0V, 1.12V and 1.26V, through the MTJ. In these voltages range, the magneto-resistance (green curve) oscillates between the two values corresponding to P and AP resistance (black reference curve). This large amplitude oscillation is characteristic of the action of the perpendicular polarizer. As shown on Fig.~\ref{fig:exp-lAR-dec}.(a), the precession is not coherent because of thermal fluctuations, the frequency is not well defined and the single-shot signals exhibit phase noise. This decoherence is responsible for the decay of the average of fifty traces, the red curves in Fig.~\ref{fig:exp-both-SS}. We also observe damped oscillations of the switching probability with the pulse width, as shown in Fig.~\ref{fig:exp-lAR-dec}.(b), with a decay due to the thermal fluctuations. The characteristic time of the decay is around 10~ns, which is consistent with the inverse linewidth observed in spin-torque oscillators with a PP ($\sim 100$~MHz)~\cite{Houssameddine2007}.\\ 

\begin{figure}[h]
\centering
\includegraphics[width=\linewidth]{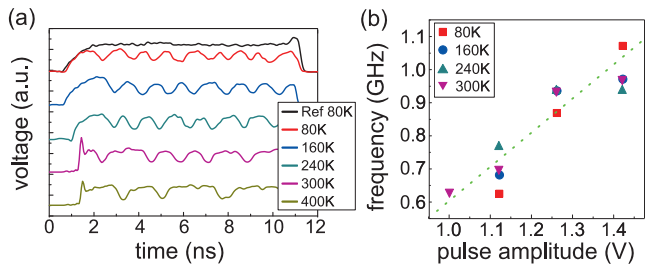}
\caption{Low aspect ratio MTJ. (a) Single-shot transmitted voltage with a pulse of 1.26 V at different temperature (from top to bottom) : 80 K, 160 K, 240 K, 300 K and 400 K. The P reference voltage is shown at 80 K (black line). (b) Oscillation frequency versus applied voltage pulse at different temperatures. The dependency is almost linear.}
\label{fig:exp-lAR-temp}
\end{figure}

Real-time measurements on a low aspect ratio sample were performed at different temperatures from 80~K to 400~K. Typical single-shot traces of the transmitted voltage at different temperatures are presented in Fig.~\ref{fig:exp-lAR-temp}.(a). They show that the precession phase and amplitude are more stable at low temperature (80~K). Due to thermal fluctuations, some precessions are missing above 240~K. The precession frequency was found to be proportional to the applied voltage of the pulse, in agreement with OPP spin-torque oscillators~\cite{Houssameddine2007}. While considering the large uncertainty on the measured frequency, the proportionality factor seems to be the same for each temperature. In the macrospin model the OPP frequency is given by~\cite{Lee2005}~:
\begin{align}
f &= \dfrac{\gamma}{2\pi\alpha}\dfrac{\hbar}{2e}\dfrac{\eta_{\operatorname{PERP}}}{M_s t} J
\end{align} 

Given that $\eta_{\operatorname{PERP}}$ and $M_s$ depend on temperature, this result seems to indicate that the thermal dependence of the STT efficiency (spin polarization) and saturation magnetization are similar which sounds reasonable. However, micromagnetic simulations show that the free layer is not uniformly magnetized in the OPP state, therefore the macrospin model is not totally adapted to describe the OPP and caution should be taken when using the formula for the frequency.

We next measured samples with a higher aspect ratio of 3.7:1, with nominal sizes of $260\times70\times3$~nm. The in-plane anisotropy field is measured to be around 10~kA/m.

\begin{figure}[h]
\centering
\includegraphics[width=0.7\linewidth]{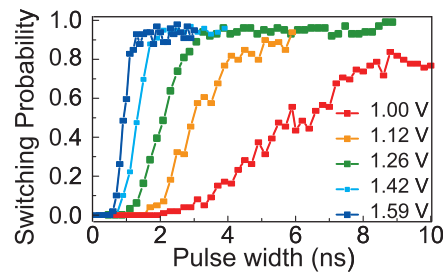}
\caption{High aspect ratio MTJ. Switching probability of an MTJ initially in P state versus applied voltage pulse width. The pulse amplitude varies from 1.0V to 1.59V.}
\label{fig:exp-hAR-prob}
\end{figure}

The real-time transmitted voltage for different applied pulse amplitude of 0.89V, 1.0V, 1.12V and 1.26V, is displayed in Fig.~\ref{fig:exp-both-SS}.(b). The MTJ is initially in the P state. As expected from the simulations, close to the critical current density, no precessional motion of the magneto-resistance was observed, only a reversal of the SL magnetization. Such time-resolved measurements with MTJ with orthogonal polarizer were never realized before and should be compared to similar switching measurements on MTJ without PP~\cite{Devolder2008}, in particular regarding the incubation time, which vanishes with the PP. When the applied pulse voltage is increased, the switching time is decreased, like with a MTJ without perpendicular polarizer. At higher voltages, hints of precessional motion start appearing. This was expected from simulations~: for a given spin polarization of the perpendicular polarizer $\eta_{\operatorname{PERP}}$, the magnetization enters into precession for high applied current density, above the current density range for reversal. This back-hopping could be reduced or even suppressed by reducing the spin polarization of the PP below the range of appearance of OPP (see Fig.~\ref{fig:macrospin} and Appendix~\ref{Annex-1}).\\
Fig.~\ref{fig:exp-hAR-prob} shows the switching probability versus pulse width for different voltage pulse amplitudes in the case of cells with large aspect ratio. In contrast to the low aspect ratio case (Fig.~\ref{fig:exp-lAR-dec}.(b)), no oscillations are observed in the switching probability. For sufficiently long pulses, the final state is fully controlled by the current direction independently of the pulse duration. Furthermore, the larger the pulse amplitude, the faster the switching. Sub-nanosecond switching was observed for pulse amplitude above 1.12V, which corresponds to a current density of $6\times10^{11}$~A/m$^2$ and a switching energy of around 1.5~pJ. This is comparable, although larger, than the values obtained previously with optimized stacks~\cite{Liu2010}.\\
\par

The integration of a perpendicular polarizer with an in-plane MTJ permitted to realize a subnanosecond bipolar reversal of the SL magnetization. For practical devices, the STT contributions from the in-plane analyzer and perpendicular polarizer must be tuned so that the PP can still provide the initial impulse which reduces the stochasticity of the switching but the final state is controlled by the current direction in the stack independently of the pulse duration. This can be achieved by increasing the aspect ratio of the cell above $\sim3$. This also improves the thermal stability of the cell. The drawback is the increase footprint of the cell but conventional CMOS SRAM have large footprint anyhow. Therefore, these high anisotropy structures with orthogonal polarizers are good candidates for realizing ultrafast MRAM for SRAM type of applications.\\

\appendix
\section{Appendix: Equilibrium stability}\label{Annex-1}
The equilibrium states of the free layer magnetization $\mathbf{m} = (m_x, m_y, m_z)$ are computed from the Landau-Lifshitz-Gilbert-Slonczewski (LLGS) equation including the STT contributions from the two polarizing layers~\cite{Lacoste2013}, the in-plane reference layer and the perpendicular polarizer. The equilibrium in-plane and out-of-plane angles are noted $\phi$ and $\theta$~:
\begin{equation*}
\begin{cases}
m_x=\sin\theta\cos\phi \\
m_y=\sin\theta\sin\phi \\
m_z=\cos\theta
\end{cases}
\end{equation*}

The equilibrium angles are solution of the LLGS equation with the time-dependent terms set to zero~:
\begin{equation}
\begin{cases}
0= - \dfrac{H_K}{2} \sin\theta \sin(2\phi) -P_z \sin\theta  - P_x \cos\theta\cos\phi  \\
0 = -\sin\theta\cos\theta (M_{\operatorname{eff}}+H_K\cos^2\phi)  + P_x \sin\phi\\
\end{cases}
\label{eq:equil}
\end{equation}
$M_{\operatorname{eff}}$ is the reduced demagnetizing field, $H_K$ is the in-plane anisotropy field and $P_x$ and $P_z$ are the spin torque amplitudes due to the reference layer (magnetized along the x-axis) and the perpendicular polarizer respectively. Their expressions are given by~:
\begin{align*}
P_{x(z)} &= \dfrac{\hbar}{2e}\dfrac{\eta_{\operatorname{LONG}(\operatorname{PERP})}}{\mu_0 M_s t}J
\end{align*}
$M_s$ is the saturation magnetization of the storage layer, $J$ the applied current density, $t$ the thickness of the storage layer, and $ \eta_{\operatorname{LONG}} $ and $ \eta_{\operatorname{PERP}} $ the STT efficiencies of the reference layer and of the perpendicular polarizer.\par

The expression of the out-of-plane angle $\theta$ at equilibrium with respect to the angle $\phi$ is computed from the second equation in eq.~\ref{eq:equil}~:
\begin{equation}
\sin(2\theta)=\dfrac{2 P_x \sin\phi}{M_{\operatorname{eff}} + H_K \cos^2\phi}
\label{eq:sintheta}
\end{equation}
For clarity, let $A=\dfrac{2 P_x \sin\phi}{M_{\operatorname{eff}} + H_K \cos^2\phi}$. We make the assumption that the demagnetizing field is dominant, so $A^2\ll 1$. From eq.~\ref{eq:sintheta}, the two possible values of the cotangent of $\theta$ are given by~:
\begin{equation*}
\cot(\theta)_{\pm}=\dfrac{1}{A}\left(1\pm\sqrt{1-A^2}\right)
\label{eq:p3c5:cottheta}
\end{equation*}
The two solutions describe an in-plane (IPS) equilibrium and an out-of-plane equilibrium (OPS), for which $\cos\theta \approx A/2$ and $\sin\theta \approx A/2$, respectively.
Replacing $\cos\theta$ and $\sin\theta$ in the first equation of eq.~\ref{eq:equil}, the expression of the in-plane angle $\phi$ at equilibrium is obtained for the IPS and OPS equilibriums~:
\begin{align}
\mbox{(OPS)~:}\quad &\cot\phi = \dfrac{-P_z}{M_{\operatorname{eff}} + H_K} \\ \mbox{(IPS)~:}\quad &\sin(2\phi) = \dfrac{-2 P_z}{ H_K + \dfrac{P_x^2}{M_{\operatorname{eff}} + H_K/2}}  
\label{eq:p3c5:phi}
\end{align}
The OPS equilibrium is always defined. However, the IPS equilibrium is defined only if the right-hand-side in the previous expression of $\phi$ is smaller than unity in absolute value, i.e.~:
\begin{align}
2 \lvert P_z \rvert< H_K + \dfrac{P_x^2}{M_{\operatorname{eff}} + H_K/2}  
\end{align}
Let $k_0=\dfrac{2e}{\hbar} \mu_0 M_s t$, from the expressions of $P_x$ and $P_z$, the criterion for the existence of an IPS equilibrium becomes~:
\begin{align}
\eta_{\operatorname{PERP}}&< \dfrac{H_K k_0}{2 J} +  \dfrac{\eta_{\operatorname{LONG}}^2 J}{2k_0 (M_{\operatorname{eff}}+H_K/2)}
\label{eq:critpx}
\end{align}
For the range of applied current densities used in applications, the second term of eq.~\ref{eq:critpx} is negligible, so the range of current densities for which the magnetization of the free layer is in OPP (because no IPS equilibrium exists and the OPS equilibrium is unstable) corresponds to current densities larger than $J_c^{\operatorname{PERP}}$.\\
It is interesting to notice that the left-hand side of eq.~\ref{eq:critpx} goes through a minimum when the current is changing. This gives rise to a maximum for $\eta_{\operatorname{PERP}}$ below which the IPS equilibrium exists for all applied current density $J$, so OPP are not expected for any current density. The maximum of $\eta_{\operatorname{PERP}}$ is given by~:
\begin{align*}
\eta^{\max}_{\operatorname{PERP}}&= \eta_{\operatorname{LONG}} \sqrt{\dfrac{H_K}{M_{\operatorname{eff}}+H_K/2}}
\end{align*}

After studying the existence of the equilibrium, one must look at their stability by linearizing the LLGS equation. After simplification with the assumption that $M_{\operatorname{eff}}$ is the dominant field, we find that, for reasonable current densities, the initial equilibrium state IPS is destabilized for applied current densities above the critical current density $J_c^{\operatorname{LONG}}$.

\par
\begin{acknowledgements}
\section*{Acknowledgements}
This work was supported by the European commission through the ERC Adv Grant HYMAGINE n$^\circ$246942.
\end{acknowledgements}

\end{document}